\documentclass[final,onefignum,onetabnum]{siamart190516}

\usepackage{amsfonts}
\usepackage{graphicx}
\usepackage{subcaption}
\usepackage{float}
\usepackage{listings}
\usepackage{color}

\definecolor{mygreen}{rgb}{0,0.6,0}
\definecolor{mygray}{rgb}{0.5,0.5,0.5}
\definecolor{mymauve}{rgb}{0.58,0,0.82}

\ifpdf
\DeclareGraphicsExtensions{.eps,.pdf,.png,.jpg}
\else
\DeclareGraphicsExtensions{.eps}
\fi

\floatstyle{ruled}
\newfloat{codelisting}{tbp}{lst}
\floatname{codelisting}{Listing}

\crefname{codelisting}{Listing}{Listings}

\headers{Towards agile test-driven development in scientific software}{T.-R. Teschner}

\title{A practical guide towards agile test-driven development for scientific software projects}

\author{Tom-Robin Teschner\thanks{Cranfield University, School of Aerospace, Transport and Manufacturing, Centre for Computational Engineering Sciences
		(\email{tom.teschner@cranfield.ac.uk}).}}

\ifpdf
\hypersetup{
  pdftitle={A practical guide towards agile test-driven development for scientific software projects},
  pdfauthor={T.-R. Teschner}
}
\fi

\begin{document}

\maketitle

\begin{abstract}
Software testing has received much attention over the last years and has reached such critical importance that agile software development practices put software testing at its core. Agile software development is successfully applied in large-scale industrial software developments but due to its granular responsibilities with roles assigned to various members of the development team, these practices may not be applicable to scientific code development, especially in an academic environment, where it is not uncommon that the codebase is developed, maintained and used by a single person. Even for collaborative scientific software development, financed through external grants, the end-users are typically still part of the development team. This is in contrast to how software is developed in many industries, where the development team and end-users are two separate entities. There are, however, many good code development practices that can be adopted for scientific software projects. Specifically, the intention of this article is to take the centrepiece of agile software development --- the test-driven development --- and tailor it to scientific and academic, single-user code development. In this study, a c++ starter project is developed and made available, based on the meson build system, which provides native support for software testing. It is used to show how a simple linear algebra application, found in many scientific and academic applications, can be developed and how simple unit, integration and system tests can be created that are managed through the meson build system. In this way, we are able to minimise software defects and reduce the risk to interpret incorrect data generated by erroneous software that may result, in the worst case, in the wrong conclusions to be drawn. Each layer of testing presents one additional layer of protection against such software defects and we will explore how these may be incorporated with minimum overhead to produce bug-free software.
\end{abstract}



\section{Introduction}
\label{sec:introduction}

\begin{figure}[t]
	\centering
	\includegraphics[scale=0.35]{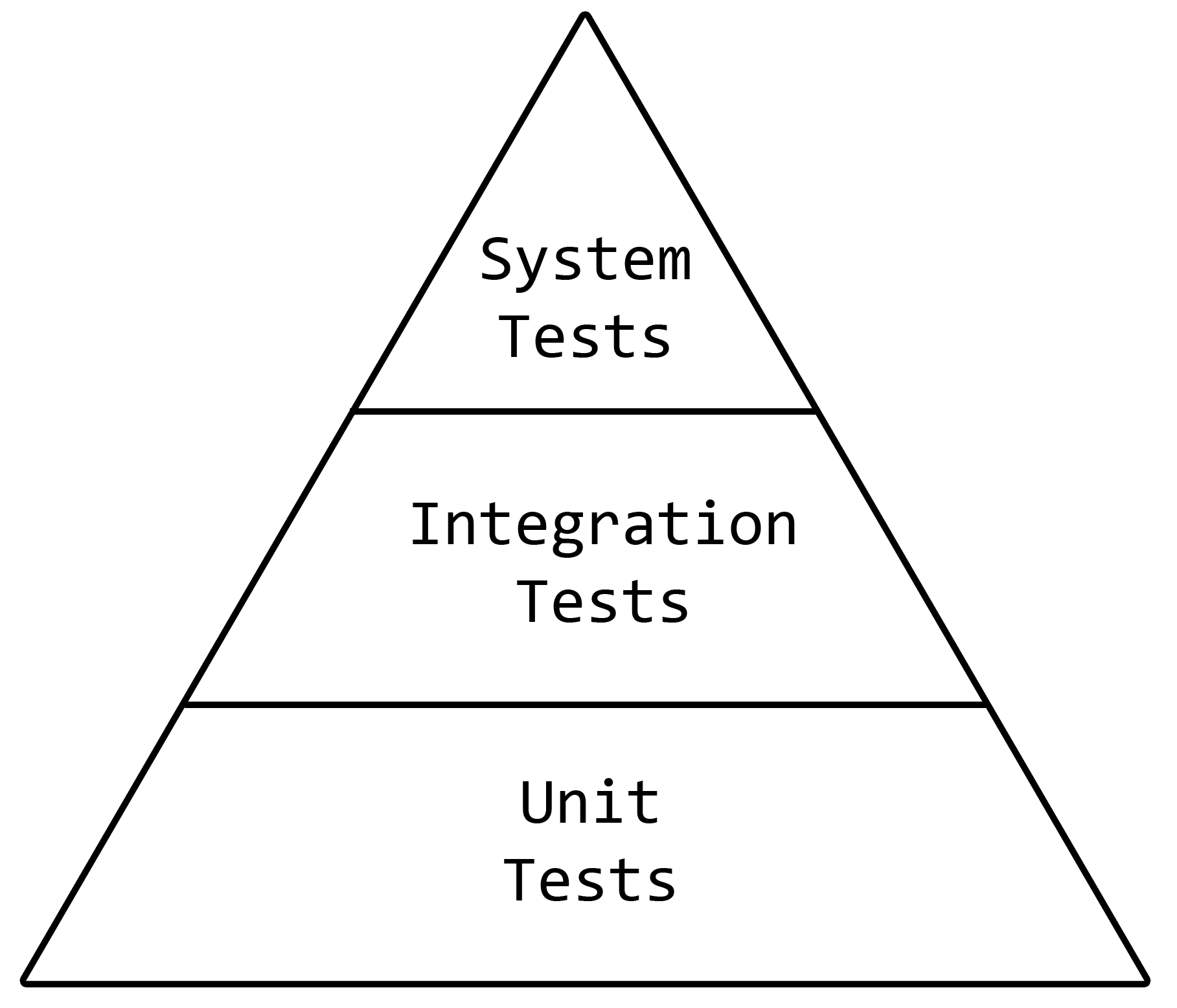}
	\caption{The testing pyramid showing that we would expect most of our tests in a test suite to comprise of unit, then integration and then system tests.}
	\label{fig:testingPyramid}
\end{figure}

Software testing is a fundamental building block of modern software engineering. Failure to maintain a clean code and a corresponding test suite may ultimately lead to disintegrated and unmaintainable code~\cite[pp. 123--124]{Martin2008}. The importance of software testing is manifested in all major software development methodologies, from the classical waterfall model through to modern day agile development strategies~\cite{Sommerville2016}. The differences in these two extremes is its agility towards the software development cycle. While the waterfall model assumes that a software plan can be developed for all stages of the development process which is then followed in sequence (with testing forming the last step), agile strategies follow an iterative development cycle, allowing for continuous code development, refactoring, review, testing, integration and release of the software. Extreme programming~\cite{Beck1999} is a popular programming strategy for agile developments, which takes good practices in programming and pushes them to the limit. This means that code is tested, reviewed and integrated on a daily basis, which can lead to weekly software releases. This approach allows for iterative end-user feedback which is then integrated into the next development cycle. In practice, this means that code is submitted on a daily basis to a web server, which will check each change of the code as soon as it is pushed to the server. Only those changes which do not break the current master version, and which have passed the code review process, will be accepted and merged with the master. This process is also known as continuous integration~\cite{Duval2007}.\\
This approach may seem to make sense for small projects with limited allocated time or software prototyping, however, software development is usually constrained by contracts, grants, external clients etc. so that agile development becomes impractical in most cases. For that reason, project management-like approaches have been developed, which are agile at its core, but allow for some form of software management and planning. A popular representative of this is Scrum. In Scrum, a so-called product backlog is produced, which contains software features and requirements that need to be developed. Then, so-called sprints are performed, typically lasting 2--4 weeks, where items from the product backlog are worked on, which are prioritised before each sprint. After a sprint is finished, work that has not been completed goes back into the product backlog, however, a sprint is never extended. After each sprint, a software increment is produced which could be released or shipped to customers. In this way, Scrum allows to have some oversight of the development and changes can continuously be integrated by adding / removing items from the product backlog.\\
During each iteration or sprint, testing takes centre stage. As this would be rather time consuming if it were to be done manually, test suites are developed that handle all of the testing automatically. These tests are the same tests executed by the continuous integration server and are available to each programmer. Following a rigid testing-orientated methodology leads to a so-called test-driven development (or TDD). In essence, TDD dictates that before any production code is written, a failing test has to be created first. This test is then added to the test suite which will automatically execute this test every time the test suite is invoked. With the failing test in place, the production code is written and refined until the test passes. At this stage, we have confidence in the code and can clean it, for example by splitting up larger functions into smaller ones or optimising algorithms. After each change to the code, we simply run the tests again, until we have cleaned the code sufficiently without breaking it.\\
The TDD is often seen as having its root in extreme programming, which was developed during the 1990s, however, earlier references can be found where the basic TDD is outlined. McCracken, for example, in 1957 wrote that "the first attack on the checkout problem may be made before coding is begun. In order to fully ascertain the accuracy of the answers, it is necessary to have a hand-calculated check case with which to compare the answers which will later be calculated by the machine."~\cite[p. 159]{McCracken1957}. In-fact, Kent Beck~\cite{Beck1999}, one of the developers and first practitioners of extreme programming states that TDD was not new but rather rediscovered and made fit for purpose to be used in extreme programming, which we have come to adopt as TDD nowadays~\cite{Beck2002}.\\
There are several layers to testing but the most common types of tests, virtually found in any application adopting a TDD are unit, integration and system tests. Unit tests focus on the smallest piece of software (a unit of code) and test them in isolation of the rest of the system. These are typically functions or methods of a class. Integration tests, on the other hand, test a larger collection of behaviour of the software, based on more than a single unit. If we follow the single-responsibility principle\footnote{The single-responsibility principle states that each class should have only one responsibility. This can be easily tested by describing the class' purpose in about 25 words. If we have to use the words ``if", ``and", ``or" or ``but" in our description, then the class has likely more than one responsibility~\cite[p. 138]{Martin2008}.}, advocated by object-orientated programming, then integration tests do typically test a whole class with all or some of its methods (or units). Alternatively, if we were to test two classes at the same time, that would also result in an integration test. System tests, on the other hand, test the whole software for its intended use. Using a linear algebra package as an example, a system test could be, for example, the solution of the linear system of equations, i.e. $ \mathbf{A}\cdot \mathbf{x} = \mathbf{b} $ through an iterative procedure, such as the conjugate gradient method. An integration test, on the other hand, could be the testing of the matrix class, by performing a matrix vector multiplication (here we have to write functionality for both the matrix and vector class that needs to be tested), which occurs during the conjugate gradient algorithm. A unit test would then be, for example, the calculation of the determinant or transpose within the matrix class. From this example we can see that we would typically expect to have more unit tests than integration tests (as we would have more units per class available for testing), but also more integration tests than system tests (as many smaller integrated components make up the full system). Therefore, the combination of unit, integration and system tests is typically described as the testing pyramid in the literature, as shown in \cref{fig:testingPyramid}.

The remaining organisation of this article is as follows: In \cref{sec:literature_review}, we review the state of testing as described in the literature followed by \cref{sec:software_testing}, where the different layers of testing are discussed and how this can be applied to scientific code development. A simple starter project containing a test suite is developed which is used in \cref{sec:case_study_developing_a_testing_suite_for_a_linear_algebra_library} to showcase how to develop a linear algebra package using a TDD approach. Selected code fragments are discussed while the full project is made available for personal study and use. \Cref{sec:conclusion} then provides a conclusion with final thoughts.

\section{Literature review}
\label{sec:literature_review}

In the following section, the literature is surveyed to highlight the main findings on software testing in general. Juristo \textit{et al.}~\cite{Juristo2006} provided an overview of the importance of testing. It is followed by Runeson~\cite{Runeson2006} who reviewed the perception of unit tests within different companies employing TDD. His analysis showed that companies echoed the advantages and disadvantages of software testing as described in classical textbooks~\cite{Sommerville2016}. Williams \textit{et al.}~\cite{Williams2009}, in particular, reflected on changing from an ad-hoc to an automated unit testing approach at Microsoft for a team of 32 people. They compared two comparable pieces of software, the first being developed without any rigid testing requirement while the second used a strict unit testing policy. While development time increased by 30\%, the overall number of defects in the software were reduced by 20.9\%. At the same time, the number of users increased by a factor of 10 while the number of defects reported by user increased only by a factor of 2.9, meaning that the average number of defects reported per user decreased. Writing additional tests requires additional code to be written, which explains the increase in production time. However, this additional development time has to be contrasted with potential extra time spent fixing the code at a later stage based on reported defects. 
Korosec and Pfarrhofer~\cite{Korosec2015} reported the transition towards an agile software development, which saw the transformation of their test suite from mostly automated system to unit tests. In-fact, system tests provide the best protection against software defects, however, they come with a high maintenance cost as described by Korosec and Pfarrhofer. When code needs to be refactored, system tests have to change which causes brittle tests. Considering that code refactoring could be exercised on a daily basis, following the extreme programming principles, automated system tests may become too much of a burden. The simple fact is that tests which are costly to maintain are tests which will eventually be turned off and thus don't provide any resistance against defects~\cite{Khorikov2020}.\\
Daka and Fraser~\cite{Daka2014} conducted a survey on unit testing and asked practitioners about their views. They found that unit tests are mainly driven by requirements and so writing new tests is seen less important than writing, fixing or refactoring production code. Furthermore, developers revealed that writing tests is not seen as an enjoyable exercise and expressed desire for more tool support for writing tests. Their study also showed that an equal number of people fixes or deletes a failing test compared to the same number of developers who try to fix the underlying error in the code. Interestingly, while most participants described that they produce tests systematically, they could not classify what constitutes a good test. In another study carried out by Gren and Antinyan~\cite{Gren2017}, 235 participants were asked about their opinion on the correlation between unit testing and code quality. Against common expectations~\cite{Martin2008,Khorikov2020}, little to no correlation was found. Similarly, Yuan and Qu~\cite{Yuan2006} found that the personal ability is holding many developers back at writing effective unit tests which highlights the responsibility each developer has to be proficient in this area to write effective test code.\\
Klammer and Kern~\cite{Klammer2015} reiterate the importance of starting a software process with a strong focus on testing, ideally implementing a TDD from the beginning. Software which is written without tests, they continue, may have a testability problem, meaning that it becomes difficult if not impossible to retrofit legacy codes with a functional and useful test suite. On the other hand, simply writing tests for the sake of it will not automatically increase the value of the production code. Tests should concentrate on the business logic and expected behaviour. Buchgeher \textit{et al.}~\cite{Buchgeher2020} investigate the way unit tests were implemented for an open source project and found that in this case, the tests were covering implementation details rather than the application's user interface. In-fact, only 17\%--34\% of the interface were covered by tests. This is a straight violation of the very definition of a unit test~\cite{Khorikov2020}, which states that unit tests should test the behaviour of the application, not the underlying implementation details. The implementation details may change, however, the interface should not. For example, when reading a text input file, a unit test focusing on implementation details would be one that checks the input file line by line and checks if the current line matches the expected line. On the other hand, a unit test focusing on the user interface is one which checks the current line that is read by the software and then tries to match that with a range of possible and expected values. In the first test, the order in which the file is processed matters (implementation detail), in the second, it doesn't. It is likely that added functionality results in changing the input file for which the first test would fail. The second test would pass but at the same time not check the newly added feature, unless we remember to add this to the test. However, following a TDD cycle, we have to write failing tests first, thus, in this case, we would add the new feature to the test and only then add the extra production code. Following this procedure means that we are not able to accidentally forget to test newly added code. Should we do so anyways for some other reason, we may pick this up by studying the coverage metric and see that a suitable test has not been written to test the new functionality. Either way, TDD helps us to protect ourselves against software defects if we follow it rigorously.\\
The aforementioned procedure to concentrate on the behaviour of the application, rather than its implementation detail, is also known as black-box testing, where no knowledge of the system under test is assumed. This is typically employed for test suites. The opposite, white-box testing, makes use of the fact that we do know the underlying implementation, which may be useful if we want to check the flow our applications takes to detect if it branches off incorrectly. Dudila and Letia~\cite{Dudila2013} proposed a hybrid model, the grey-box testing, which takes advantages of both approaches.\\
Trautsch and Grabowski~\cite{Trautsch2017} analysed how unit testing is used in open source projects and concluded that most developers of those projects believe to write more unit tests than they actually do and that most projects feature an insufficient amount of tests in general. They investigated the cause for that by studying the commit history of the different codes and found four distinct patterns to describe these open source codes. Either there was in initial high number of tests which then either stagnated or even dropped, or the number of tests were constantly increasing or constantly low. One explanation could be that unit test require additional time to create or may be seen as an obstacle, especially if the underlying production code is difficult to test. Another reason coming from the extreme programming methodology is that every piece of code is peer-reviewed and thus takes up additional time. To reduce time spent on fixing common issues with unit tests, Ramler \textit{et al.}~\cite{Ramler2016} presented a methodology of static unit testing. Here, the code of the unit tests is automatically (statically) analysed to reveal syntax errors, violation of the testing pattern and antipattern that should be avoided during testing. This may help to concentrate on the logic of the test rather than the structure of it, similar to static code analysing tools that can highlight potential problems with the syntax of the code.
Another automated tool, called OUTFIT, was created by Holling \textit{et al.}~\cite{Holling2016} which creates integration tests automatically, based on unit tests with a high coverage metric. They applied their approach to an engine control system and they showed that this automated procedure has the potential to flag unused and superfluous code fragments. Sun \textit{et al.}~\cite{Sun2019} concentrated on an automatic toolchain covering Gcov (coverage tool), Jenkins (continuous integration server) and QTest (testing tool) and showed how to automate the testing procedure.\\
The preceding studies have used the term test coverage, which indicates how much of the production code is actually tested, expressed as a percentage. Early adopters of TDD were striving for a 100\% test coverage metric which may seem useful at first. However, this means that a lot of tests have to be written for code fragments for which there is little value. Khorikov~\cite{Khorikov2020} advocates that unit tests should only test core business logic and nothing else. This is supported by Antinyan and Staron~\cite{Antinyan2019}, who investigated the number of software defect as a function of test coverage for a code consisting of $ 2 \cdot 10^6 $ lines. They showed that only 9\% of defects could be explained by the test coverage metric, while the remaining 91\% of defects were invariant to the code coverage metric. A useful way, however, to test the boundaries of the application and test near regressions extensively is through parametrised test, where the same test is executed several times with different input parameters. This may help to detect issues for certain combinations of parameters~\cite{Tillmann2010,Xie2016}.\\
\\
As mentioned in the introduction, we strive for a testing pyramid with unit tests at the bottom (the majority of tests), followed by integration and only then by system tests. Code analysis performed by Contan \textit{et al.}~\cite{Contan2018} on five software projects using an agile development process revealed, however, that those software projects did not show any testing pyramid structure. They further argue that there is no scientific evidence and references to support the testing pyramid idea. It may, however, be entirely possible that the TDD methodology was not rigorously or inconsistently enforced, which can very easily lead to degenerated testing pyramid shapes. However, while unit tests can be clearly defined, as done in \cref{sec:testing_in_scientific_applications}, integration tests are defined as anything a unit and system test isn't. This may not be a very helpful definition and so it is conceivable that this may lead to confusion over how to separate integration tests from unit tests. Another approach as advocated by Shore~\cite{Shore2004} is the fail fast principle, which can either replace or at least complement integration tests. The fail fast principle states that instead of trying to recover software from an unrecoverable state through exception handling, the code should fail immediately when it finds that it is no longer safe to continue. The benefit is that from experience we know that trying to recover software after it has thrown an exception, it may lead to a crash further down the road. The problem now is that the reason for the program to crash is seemingly unrelated to the root cause and it will take extra debugging time to figure out exactly where the program went wrong. This is a time-consuming exercise and so Shore suggests that failing should be done when an error occurs. Some criticism stated that this will lead to software constantly crashing, however, Shore argues that the opposite is the case, as software defects are quickly found (either during development or user acceptance testing) and thus can be easily fixed. In-fact, a combination of fail fast and integration tests provides possibly the best protection against regressions and should be used in conjunction.

\section{Software testing}
\label{sec:software_testing}

The following section provides a discussion of how software testing works in practice and how the TDD can be adopted for scientific software projects, especially developed in an academic environment.

\subsection{Test-drive development tailored to scientific applications}
\label{sec:testing_in_scientific_applications}

We have discussed the software testing pyramid consisting of unit, integration and system test, shown in \cref{fig:testingPyramid}, which is a result of agile TDD. The opposite can also sometimes be found and is referred to by some as the ice cone shape, with the majority of tests being system tests with little or no integration and unit tests. Since system tests require the manual inspection of test results (and potentially take a long time to execute), they cannot be used for strict TDD as we want to run tests frequently and fast. This type of testing is commonly employed by scientific software projects. There are, however, many more layers to software testing and a thorough review is given by Sommerville~\cite{Sommerville2016}. For example, we can further introduce release and user testing, where release testing is defined as a process which should not involve the development team but rather an outside person to ensure that the software is meeting the requirements. User testing, on the other hand, involves the end-user directly and this step can be broken down into alpha, beta and acceptance testing. During alpha testing, only a few selected candidates are given access to the software, which are working closely with the development team and provide direct feedback. During beta testing, a release is made to a larger group of people who may discover any software bugs which they can raise with the development team. Acceptance tests require the end-user to interact with the system and decide whether the software is fit for purpose.\\
\\
There is a subtle but significant difference between general purpose and scientific software development, especially in an academic environment. General purpose or industrial software development assumes that a piece of software is commissioned and worked on by a team of software engineers which are, however, not part of the end-user group. In an academic environment, however, the software developer is most commonly also the end-user (and in most cases also the only one). Furthermore, software engineers work full time and in a team on the same codebase, while an academic member of staff may not be able to justify full time code development. For these reasons, the potential gains in productivity by adopting an agile development process may not justify the additional time required for the additional project management. Therefore, an agile development process may not be lucrative for academic code development. As highlighted in~\cite{Sommerville2016}, agile development is not suited for all projects. However, there are many good practices that we as scientists can adopt from these development methodologies. Specifically, this article focuses on the integration of a TDD cycle into scientific software projects. This does not only have the advantage of reducing regressions (bugs) in scientific software, adopting a TDD cycle forces us to write code which is testable. Testable code typically produces clean code which in turn is easy to maintain. An excellent discussion on the importance of clean code can be found in~\cite{Martin2008}, while a very thorough review of best practices in testing can be found in~\cite{Khorikov2020}.\\
\\
In order to establish how TDD in a scientific or academic environment could look like, we need to classify the scenarios under which code is developed. The two most common ones are described below and will be the focus hereafter.\\

\begin{enumerate}
	\item In the first scenario, the code is developed by a single person, for example to test a newly developed model or hypothesis. This may be part of a publication or proof of concept study. In either case, the software developer is also the end-user at the same time.\\
	\item In the second scenario, the code is developed by two or more people. This can be either as part of a pan-centre activity (including collaboration with supervised students) or a wider collaboration with other departments and / or external contributors. Larger teams may be working on a piece of software as part of a research project with external funding. While in a classical software testing approach the end-user would be identical to the body providing the funding, for research projects the end-user is still typically part of the development team.\\ 
\end{enumerate}
Thus, we can see that regardless of the nature, we as developers are also the end-user and thus testing will fall entirely into our domain. This increases the testing requirements on ourselves, however, this also means that we can combine all stages of user testing into a single one, without differentiating between alpha, beta and acceptance testing.\\
One may be tempted to include open source projects in the list of scenarios above. However, given that there is usually a large group of people simply using the software (without developing it), a clear separation between developer and end-user is given and a classical software development approach can be used. It is up to the project maintainer, though, to ensure that user contributions are tested before merged into the production codebase.\\
\\
How does this affect the software development cycle for scientific and academic software? First of all, we should be following a TDD approach, consisting of unit, integration and system tests. Doing so will provide us with the best protection against software defects while providing confidence during code refactoring that everything is still working as intended. We do not, however, require any further testing than that. Specifically, user and system testing are synonymous. System tests are the responsibility of the developer while user testing should be done by the end-user. As these roles are occupied by the same person, we can safely merge them into a single entity. This reduces the testing requirements but highlights that we still need to provide unit, integration and system tests. The following section will provide an overview of how they are defined and how we would typically arrange them.

\subsection{Unit, Integration and System tests}
\label{sec:tests_used_during_scientific_testing}

What then constitutes a good unit test? According to Khorikov~\cite[p. 21]{Khorikov2020} there are three core principles that define a good unit test. These define a unit test as something that:\\
\begin{itemize}
	\item verifies a small piece of code (also known as a \textit{unit})
	\item does it quickly
	\item does it in an isolated manner\\ 
\end{itemize}

Let's examine these in parts. The first item may seem trivial but it should be stressed here that the name unit in unit test suggest that the smallest possible unit within a code is to be tested. The smallest unit of code are typically functions or methods within classes. Ideally, these functions or methods usually contain several lines but depending on the design philosophy they could contain hundreds of lines. As these are going to be the smallest unit within our code, they should probably not contain hundreds of lines. Martin~\cite[p. 34]{Martin2008} suggest an upper limit of 4 lines of code per function and more importantly, that they should only do one thing. The upper limit may seem restrictive and sometimes it is, especially if we want to incorporate the fail fast principles~\cite{Shore2004}, which needs to check the state of the program within the function or method. However, even in those cases one should probably try to stay below 10 lines of code per function. If the function still contains more lines, even though it is only doing one thing, this may highlight issues with the general code design, for example, the function needs to set up dependencies itself (which really should be done within a constructor) or it may hint a lack of \textit{encapsulation}. Either way, trying to make our functions and methods as small as possible forces us to produce clean code which is self-documenting and easy to follow. All of this is a result of the first item in the above list, trying to verify a small piece of code which should be a function or method with a single responsibility.\\
The second item states that it should be done quickly. This requirement may not seem to be immediately obvious, however, remember that in TDD we want to run all tests frequently to ensure that our current implementation is not breaking any existing code. The only way to do that is to run tests which are fast. If they take too long, the tests will simply not be run and this defeats the entire purpose of writing tests in the first place. They are there to help us catch software defects and that they can do only if we run the test suite frequently. Thus, a single unit test should probably run within milliseconds.\\
The third item requires the test to be run in an isolated manner. This is typically a requirement that no external dependencies should be used. External dependencies could be web servers that are queried for information or a database in live use containing program critical information (typically containing user information but in the context of scientific applications we may want to have a database storing results for later evaluation, such as user data and their answers given during an experiment). Databases can be slow to query or may not be available for testing, as we do not want to pollute the real database with test data. Furthermore, as testing should be done frequently, it may not be possible to access the real database with high frequency. Additionally, there may be some web-services which the application depends on but to which a limited number of requests can be made per day. In all of these cases we are not able to interact with the real environment and have to somehow replace it with a fake environment. This is commonly referred to as mocking. This approach replaces any external dependencies by mocks which mimic the interaction with that now unavailable dependency. For example, we may have a database which has certain functionality (such as storing user data from an experiment) but instead of storing fake test data in the actual experimental database, we could mock that database and ensure that any call a function is making to said database is being caught by the mock. If a return value is required by that function, we can instruct the mock to return a specific value based on the called function and in this way, we have completely removed the external dependency.\\
\\
An integration test, then, is anything which violates any of the three core principles of a unit test. For example, if two units are tested together rather than in isolation, we would have an integration test. Or, if we decided to test a time-consuming component which should not run with all other fast executing unit tests, we would classify that as well as an integration test.\\
A system test, sometimes also referred to as an end-to-end test, is a subset of integration tests~\cite{Khorikov2020} and typically tests the whole system under real working conditions, including any real dependencies without mocking them.\\
\\
This brings us to the anatomy of a unit test. How should it be structured? First of all, there are many useful and well-maintained testing frameworks that can help with many of the aforementioned task that a unit, integration and system test should fulfil. For example, while it is possible to write custom mocks whenever they are needed (in this case they may also be referred to as a spy), one may want to explore a testing framework which already has this functionality inbuilt. Even for simpler cases, a testing framework can help with many seemingly simple tasks. One of them is the comparison of floating point numbers. This seems trivial but when we want to compare two numbers, rounding errors due to single- or double-precision will very likely influence the outcome of the test. For this reason, we may wish to either use a framework that takes care of that matter for us or at the very least write our own helper functions that compares floating point values. An example in c++ is given in \cref{lst:compare_floating_point_numbers}.

\begin{codelisting}
\begin{lstlisting}
#include <array>
#include <limits>
#include <cmath>
#include <cassert>

template<
	typename T,
	typename std::enable_if_t<std::is_floating_point<T>::value>* = nullptr
>
void assert_floating_point_eq(
	const T &firstValueToCompare,
	const T &secondValueToCompare,
	const T &tolerance = 1
) {
	const T differenceBetweenValues = std::fabs(firstValueToCompare - secondValueToCompare);
	const std::array<T, 2> absoulteValues{std::fabs(firstValueToCompare), std::fabs(secondValueToCompare)};
	const T largestAbsoluteValue = absoulteValues[1] > absoulteValues[0] ? absoulteValues[1] : absoulteValues[0];
	const bool isEqual = differenceBetweenValues <= largestAbsoluteValue * std::numeric_limits<T>::epsilon() * tolerance ? true : false;
	assert(isEqual && "floating point values don't match");
}
\end{lstlisting}
\caption{Example function to compare floating point numbers.}
\label{lst:compare_floating_point_numbers}
\end{codelisting}

The template keyword on line 6--9 ensures that we can operate on any floating point type, however, if we were to try to pass in a type which is not a floating point type (for example, an integer), the construct provided will throw an error and say that the function is not defined. This form of only defining a function for certain types (for which they make sense) while disabling them for other types (for which the function doesn't make sense) is referred to as ``Substitution failure is not an error", or SFINAE for short. Note here that with the introduction of \textit{concepts} in the c++20 standard we may rewrite this template definition using a more expressive syntax.\\
The actual body of the function is comparing floating point numbers relative to each other and not their absolute values. A very well documented explanation of that can be found in~\cite{Dawson2012}, where inspiration for the given code has been taken from. An alternative approach to the above relative comparison would be the usage of the \texttt{std::nextafter()} functionality, provided by the c++ standard template library, which returns the next possible value that can be represented given the current floating point type (i.e. single- or double-precision). Either way, the function given in \cref{lst:compare_floating_point_numbers} may be used as shown in \cref{lst:use_floating_point_numbers_comparison}

\begin{codelisting}
\begin{lstlisting}
int main() {
	// OK, same value
	assert_floating_point_eq(3.14, 3.14);
	
	// OK, same value
	assert_floating_point_eq(0.123456789, 0.123456789);
	
	// OK, although not the same value, it is precise enough for single precision
	assert_floating_point_eq(0.123456789f, 0.123456780f);
	
	// not OK, values are not the same
	assert_floating_point_eq(3.14, 3.15);
	
	// not OK, function is not defined for integer values
	assert_floating_point_eq(3, 3);
	return 0;
}
\end{lstlisting}
\caption{Example test making use of the function defined in \cref{lst:compare_floating_point_numbers}.}
\label{lst:use_floating_point_numbers_comparison}
\end{codelisting}

We are now in a position to write unit tests, and it is a good practice to follow the AAA principle which stand for Arrange, Act, Assert~\cite{Khorikov2020}. This may be best illustrated using an example. Suppose we want to write a custom class to handle complex numbers. Furthermore, assume that our current use is simply to add complex numbers in our application and to get their respective real and imaginary part. Before we write any production code to handle this logic, we would first think about how such a functionality can be tested and write a failing test, as dictated by TDD. A failing test following the AAA principle may be written as shown in \cref{lst:unit_test_complex_numbers}.

\begin{codelisting}
\begin{lstlisting}
#include <cassert>

int main() {
	// Arrange
	ComplexNumber<int> c1(3, 7);
	ComplexNumber<int> c2(-2, 1);
	
	// Act
	const auto result = c1 + c2;
	
	// Assert
	assert(result.getRealPart() == 1);
	assert(result.getImaginaryPart() == 8);
	
	return 0;
}
\end{lstlisting}
\caption{An example unit test for complex number addition following the AAA principle.}
\label{lst:unit_test_complex_numbers}
\end{codelisting}

We can see that during the arrange phase, we simply set up the problem and all that is necessary to test the underlying unit of code, in this case we want to test the addition of two complex numbers and thus set up our project to have such two different numbers. Note that we intend to use templates here and we may as well want to test floating point numbers, for which case the \texttt{assert\_floating\_point\_eq()} function defined in \cref{lst:compare_floating_point_numbers} may be used.\\
During the act phase, we should have exactly one statement. If we have more than one statement here, we are not testing a single piece of code anymore and thus would classify that as an integration test. Here, we test the addition of the two previously defined complex numbers.\\
The final step is the assert phase, where according to Khorikov~\cite{Khorikov2020} you should have a single assert. Martin~\cite{Martin2008}, on the other hand, states that you should have a single \textit{behavioural} assert. In this case, to ensure the correct functionality, we would have two assert statements, one for the imaginary and another for the real part of the complex number. Both asserts check a single behaviour and thus still pass as a unit test.\\
Now that the test is written, if we compile the test, we would of course get an error from the compiler stating that the \texttt{ComplexNumber} class is not defined. We would write that next and an example of that class could look like the one provided in \cref{lst:complex_numbers_class}

\begin{codelisting}
\begin{lstlisting}
template<typename Type>
class ComplexNumber {
public:
	ComplexNumber(const Type &real, const Type &imaginary)
	: real_(real), imaginary_(imaginary) { }
	
	Type getRealPart() const { return real_; }
	Type getImaginaryPart() const { return imaginary_; }

	ComplexNumber operator+(const ComplexNumber &other) {
		ComplexNumber c(0, 0);
		c.real_ = real_ + other.real_;
		c.imaginary_ = imaginary_ + other.imaginary_;
		return c;
	}

private:
	Type real_;
	Type imaginary_;
};
\end{lstlisting}
\caption{Implementation of the complex number class to make the unit test shown in \cref{lst:unit_test_complex_numbers} pass.}
\label{lst:complex_numbers_class}
\end{codelisting}

If we compile the test again with the given functionality, it will pass and we can safely use the functionality in our production code. We have thus established how to write a simple unit test and in-fact all unit test should follow the same AAA principle, which cleanly separates them from integration and system tests while improving readability, as they produce short test bodies with an expected structure.

\subsection{Setting up a project for testing}
\label{sec:setting_up_a_project_for_testing}

Apart from a conceptual understanding of what should go into a test suite, it is important to select the right tools that allow for direct integration into the developer's workflow. There are plenty of testing suites available covering a range of programming languages, however, this section is aimed at providing a minimal setup that allows for a straight-forward test suite integration without having to manage external dependencies.\\
\\
The foundation to this problem will be the meson and ninja build system (\url{https://mesonbuild.com/} and \url{https://ninja-build.org/}). These two tools are readily available through either the operating system's or python's package manager. In order to visualise the code coverage, we will also be using LCOV which, again, can be installed through the system's package manager within a UNIX environment. There are, of course, a plethora of other build systems available with the same functionality. Make is in common use and may be favoured by some if not many over learning a new build system. The advantage with meson is its syntax and design philosophy; it knows exactly what it wants to be and does it exceptionally well. Once the syntax is learned, writing a build file becomes just as natural as writing the corresponding source file. Meson largely follows python syntax but it is not a Turing complete language, for which good reasons exist. Thus, since python features a relatively user-friendly syntax (while also being a very popular language), most people will not have difficulties adopting meson as a build system. Ninja is similar to Make and is required by meson. In-fact, meson only produces the final ninja files which are required to build a project and hence can be regarded as a pre-processor for ninja. The reason, then, to use meson instead of ninja directly is that meson takes care of all additional steps under the hood which means that we as developer have to spend less time writing build scripts and more time writing productive code. As an example, meson defines targets such as \texttt{executable()}, \texttt{static\_library()} and \texttt{shared\_library()}, which all require as input arguments a name and a list of source files. All other compiler and linker flags as well as the compilation process itself is removed from the developer and we can influence them through an expressive configuration interface (for example, by specifying the target to be a debug or release build, which will set appropriate optimisation flags, among other things). Furthermore, it has build system specific capabilities ninja doesn't (for example, generating source files from user-specified configuration data) which makes it a prime candidate to manage the entire build process. The only thing we as developer need to know is how to invoke ninja, but a deeper understanding of it is not required. Strictly speaking, LCOV is also not required, but it allows us to visualise the test coverage which helps in identifying if additional tests are required or if the codebase is sufficiently tested.

\begin{figure}[tp]
	\centering
	\includegraphics{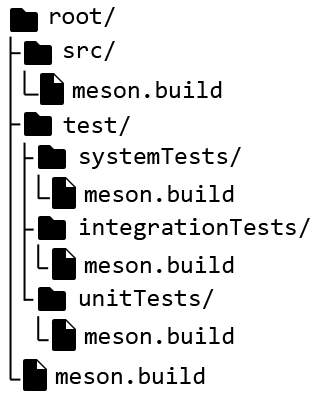}
	\caption{Basic folder structure of a starter project using the meson build system.}
	\label{fig:folderStructure}
\end{figure}

The basic folder structure of a starter project could look like the one shown in \cref{fig:folderStructure}. Inside the root directory, there are two main folders, one containing the source files and another containing all test files. Each sub-folder is equipped with their own build script which helps to separate target specific build commands. For example, the build file withing the source folder only deals with source files and their target within that folder, while it does not require any knowledge about the tests. Inside the test folder, we simply make use of any target defined within the source folder. We also see that we have split all tests into unit, integration and system test. This is a subjective choice which orders each test into its respective category, which forces us to decide beforehand what type of test we are going to write. More commonly, however, a single test folder strategy is adopted and no differentiation between these three types are made. There is no right or wrong here, but for the sake of clarity, we adopt a segregated approach.

\begin{codelisting}
\begin{lstlisting}[language=python]
# root/meson.build

# basic project settings, used each time when we configure a project
project(
	'nameOfSoftware',
	'cpp',
	default_options: ['cpp_std=c++14', 'b_coverage=true'],
	version: '0.0.0',
	license: 'MIT'
)

# handle source files and their build target in sub-directory
subdir('src/')

# handle tests and their targets in sub-directory
subdir('tests/unitTests')
subdir('tests/integrationTests')
subdir('tests/systemTests')
\end{lstlisting}
\caption{Structure of the \texttt{meson.build} file in the root directory.}
\label{lst:meson.build_root}
\end{codelisting}

Next, we examine the build scripts in turn. In the root directory of the project, we define the \texttt{meson.build} as shown in \cref{lst:meson.build_root}. A few explanations are in order here, first of all, there can only be a single \texttt{project} definition, which is typically located as the first thing within the root folder. It contains two mandatory properties, which are the name of the software and the language (here c++), while a number of optional arguments can be given. In this case, for example, we specify the version and license, along with default options. These default options are used every time a project is configured. Here we state that we want to use the c++ 2014 standard and that we want to enable coverage tracking, so we know how much of our code is actually covered by tests (as a percentage). After we have set up the project, we change into each sub-directory containing a \texttt{meson.build} file which will execute specific tasks for this sub-directory.

\begin{codelisting}
\begin{lstlisting}[language=python]
# root/src/meson.build

# link back to the root directory
root = '../'

# add source files
cppSourceFiles = [
	'sourceFile1.cpp',
	'sourceFile2.cpp',
	'sourceFile3.cpp'
]

# create shared library of all cppSourceFiles which allows to easily include them in tests
nameOfLibrary_lib = shared_library('nameOfLibrary', cppSourceFiles, include_directories: root)
\end{lstlisting}
\caption{Structure of the \texttt{meson.build} file in the \texttt{src/} directory.}
\label{lst:meson.build_src}
\end{codelisting}

The build script within the source folder is shown in \cref{lst:meson.build_src}. In this case, we collect all source files within a list and pass that list as an argument to the \texttt{shared\_library()} function. This function, again, requires two arguments, the name of the library and the corresponding source files to compile into a shared library. We could have also compiled it into a static library simply by changing the function to \texttt{static\_library()} as mentioned above.

\begin{codelisting}
\begin{lstlisting}[language=python]
# root/test/unitTests/meson.build

# link back to the root directory
root = '../../'

sourceFile1Test = executable(
	'sourceFile1Test',
	'sourceFile1Test.cpp',
	link_with: nameOfLibrary_lib,
	include_directories: root
)
test('unit test: test source file 1', sourceFile1Test)
\end{lstlisting}
\caption{Structure of the \texttt{meson.build} file in the \texttt{test/unitTests} directory.}
\label{lst:meson.build_test_unitTest}
\end{codelisting}

An example build script from the unit test sub-directory could take the form as shown in \cref{lst:meson.build_test_unitTest}.
Here we see that we need to specify an executable as we want to run our tests. We give it again the two mandatory arguments, i.e. a name and a source file (assuming here that a source file \texttt{sourceFile1Test.cpp} is available within the same folder as the \texttt{meson.build} file) but also two additional arguments. One of them, \texttt{link\_with} expects any form of library that was created as part of this project. We have generated a shared library inside the source sub-directory which we can use in the tests. A variable once defined will be globally seen by any build script. Hence, we can use the \texttt{nameOfLibrary\_lib} created before and link it with the test executable. Note that the syntax will not change if we specify the library to be static instead. The reason we do it this way is that we want to test single units of code from the source files and by compiling them into a library we can test everything individually. Alternatively, we could also include each source file within the corresponding test, but that approach may be too labour intensive and simply linking against a source file library may be easier. The reason we include the root here as a directory (and in the previous build scripts), is that we can declare include directives within the source files relative to the root. This is not required but yields more readable include directives. For example, if we wanted to include a header file \texttt{headerFile.hpp} from within the \texttt{src/} sub-directory inside our \texttt{sourceFile1Test.cpp} file, we would need to include it using \texttt{\#include "../../src/headerFile.hpp"}. However, including the root in the executable definition allows us to simply write \texttt{\#include "src/headerFile.hpp"}.\\
The final step in the build script is the \texttt{test()} function, which calls the executable with a string identifying the test. Moreover, every time we call the test suite, every executable within a \texttt{test()} function will be executed, which makes it easy to add test to the suite. Once we run the test suite, every test will be executed and at the end we will be given a summary of all tests that passed and failed.\\
\\
Finally, if we have set up our project with the above data structure and build files, we need to configure our project. On a UNIX machine, one would type \texttt{meson build} into a terminal (within the root folder) to create a folder called \texttt{build/}, which is used to store any output generated by meson (executables, libraries, object files, logs, etc.). This step only needs to be done once. Next, we would use ninja to generate any output. If we simple want to run the test suite, we would type \texttt{ninja -C build test}, where the \texttt{-C build} command instructs ninja to run inside the build folder (where the ninja build script will be stored by meson). The \texttt{test} command instructs meson to run the test suite and is a default name given by meson. If we don't specify any argument, then ninja will run all targets found in the project.\\
If we have specified our project to include coverage as well, we can run \texttt{ninja -C build coverage} to generate a coverage report and output it as html files. With these few commands we are now able to populate the \texttt{src/} and \texttt{test/} folder with content, which we will do in the next section.

\section{Case study: Developing a testing suite for a linear algebra library}
\label{sec:case_study_developing_a_testing_suite_for_a_linear_algebra_library}

In order to demonstrate how a TDD cycle can be integrated into scientific or academic software development, a case study is presented in which a linear algebra library is created to solve a linear system of equations. The full source code is made available online~\cite{Teschner2020} of which selected parts are discussed in the following.\\
Let us assume that the requirement here is that a simple 1D heat equation should be solved implicitly. We would like to do so using the conjugate gradient (CG) method. The algorithm for the CG method, iteratively solving the linear system of equation $ \mathbf{A}\mathbf{\varphi} = \mathbf{b} $, is given by Eqs.\cref{eq:CG_start}--\cref{eq:CG_stop}.

\begin{align}
\mathbf{r}^0 = \mathbf{d}^0 &= \mathbf{b} - \mathbf{A} \mathbf{\varphi}^0 \label{eq:CG_start} \\
\alpha &= \frac{(\mathbf{d}^{n})^T\mathbf{r}^n}{(\mathbf{d}^n)^T\mathbf{A}\mathbf{d}^n} \\
\mathbf{\varphi}^{n+1} &= \mathbf{\varphi}^n + \alpha \mathbf{d}^n \\
\mathbf{r}^{n+1} &= \mathbf{r}^n - \alpha \mathbf{A} \mathbf{d}^n \\
\beta &= \frac{(\mathbf{r}^{n+1})^T\mathbf{r}^{n+1}}{(\mathbf{r}^n)^T\mathbf{r}^n} \\
\mathbf{d}^{n+1} &= \mathbf{r}^{n+1} + \beta \mathbf{d}^n \label{eq:CG_stop}
\end{align} 

In order to implement that algorithm, we need to break it down into a matrix and vector class. Each class should define appropriate operator overloads and these we can test individually. Specifically, we need to be able to perform the following operations:\\

\begin{enumerate}
	\item scalar $ \cdot $ vector
	\item vector transpose $ \cdot $ vector
	\item vector + vector
	\item vector - vector
	\item scalar $ \cdot $ matrix
	\item matrix $ \cdot $ vector\\
\end{enumerate}

\begin{codelisting}
\begin{lstlisting}
#include <cassert>
#include <vector>

#include "src/vector.hpp"

int main() {
	// Arrange
	LinearAlgebra::Vector vector({1, 2, 3});
	const double scaleFactor = 2;
	
	// Act
	const auto scaledVector = scaleFactor * vector;
	
	// Assert
	assert(scaledVector(0) == 2);
	assert(scaledVector(1) == 4);
	assert(scaledVector(2) == 6);
	
	return 0;
}
\end{lstlisting}
\caption{Unit test for the scalar $ \cdot $ vector multiplication.}
\label{lst:unit_test_scalar_vector_multiplication}
\end{codelisting}

As an example, we may write the scalar $ \cdot $ vector test in the way shown in \cref{lst:unit_test_scalar_vector_multiplication} and store it inside the \texttt{test/unitTests/} directory. We have not yet defined the \texttt{src/vector.hpp} file and thus trying to compile the test will result in a compilation error. We may provide the vector class next and only add functionality that is required to pass the test. This is shown in \cref{lst:vector_class_definition_and_implementation}.

\begin{codelisting}
\begin{lstlisting}
#ifndef VECTOR_HPP
#define VECTOR_HPP

#include <iostream>
#include <vector>

namespace LinearAlgebra {

class Vector {
public:
	using VectorType = std::vector<double>;

public:
	Vector(const VectorType &inputVector);
	Vector &operator*(const double &scaleFactor);
	friend Vector &operator*(const double &scaleFactor, Vector vector);

private:
	VectorType vector_;
};

Vector::Vector(const Vector::VectorType &inputVector) : vector_(inputVector) { }

Vector &Vector::operator*(const double &scaleFactor) {
	for (auto &vectorComponent : vector_)
		vectorComponent *= scaleFactor;
	return *this;
}

Vector &operator*(const double &scaleFactor, Vector vector) {
	return vector * scaleFactor;
}

} // namespace LinearAlgebra

#endif
\end{lstlisting}
\caption{Vector class definition required to make the unit test shwon in \cref{lst:unit_test_scalar_vector_multiplication} pass.}
\label{lst:vector_class_definition_and_implementation}
\end{codelisting}

Note that we define both the \textit{scalar} $ \cdot $ \textit{vector} and \textit{vector} $ \cdot $ \textit{scalar} case here. Since the first case defines the operator overloading for the type of the scalar (here double), which is independent of the vector class, we have to define it as a \textit{friend} of the vector class. Once these files are added, we can invoke first \texttt{meson build} and then \texttt{ninja -C build test}, assuming we have set-up the meson build scripts as explained in \cref{sec:setting_up_a_project_for_testing}, which will then print the message shown in \cref{lst:ninja_output_single_unit_test} to screen.

\begin{codelisting}
\begin{lstlisting}
Found ninja-1.10.0 at /usr/bin/ninja
[0/1] Running all tests.

1/1 unit test: scalar vector multiplication OK             0.01s

Ok:                 1 
Expected Fail:      0   
Fail:               0   
Unexpected Pass:    0   
Skipped:            0   
Timeout:            0  
\end{lstlisting}
\caption{Output after running \texttt{ninja -C build test} with a single unit test defined.}
\label{lst:ninja_output_single_unit_test}
\end{codelisting}

We see that the test has passed within 10 milliseconds which allows us to proceed with our implementation. It is worth highlighting here, that the reason the test has passed is because it reached the end of the main function, i.e. the program returned 0 which is seen by meson as a successful pass. Returning any other value will be interpreted as a test failure. If the assert in the test would have failed, that would have been picked up by meson as well as a test failure and would have been reported that way, along with an error message (in this case the line where the test failed).

\begin{codelisting}
\begin{lstlisting}
#include <cassert>
#include <vector>

#include "src/matrix.hpp"
#include "src/vector.hpp"

int main() {
	// Arrange
	LinearAlgebra::Matrix matrix({{1, 2, 3}, {4, 5, 6}, {7, 8, 9}});
	LinearAlgebra::Vector vector({6, -2, 6});
	
	// Act
	const auto resultVector = matrix * vector;
	
	// Assert
	assert(resultVector(0) == 20);
	assert(resultVector(1) == 50);
	assert(resultVector(2) == 80);
	
	return 0;
}
\end{lstlisting}
\caption{Integration test for the matrix vector multiplication.}
\label{lst:integration_test_matrix_vector_multiplication}
\end{codelisting}

In a similar way, we can provide unit tests for other vector operator and also define a matrix class. However, it is worthwhile to investigate the case of the matrix vector multiplication. Assume that the matrix and vector class are defined but the implementation for their product is not yet available. We may write a test as shown in \cref{lst:integration_test_matrix_vector_multiplication}. Next we would need to provide the implementation. We have two options here, either we can specify that inside the matrix or vector class. As the product returns a vector, the implementation is done within the vector class and shown in \cref{lst:matrix_vector_multiplication_implementation}. As we require both the vector and matrix class here (and their implementation is important to the outcome of the calculation), we would classify this test as an integration rather than a unit test. In a similar way, we may set up a test for the conjugate gradient method, consisting of the vector, matrix and the conjugate gradient class itself. This will produce another integration test.\\

\begin{codelisting}
\begin{lstlisting}
\\ the following is added to the vector class
friend Vector operator*(const Matrix &matrix, const Vector &vector);

\\ the following is added outside the vector class
Vector operator*(const Matrix &matrix, const Vector &vector) {
	assert(matrix.getNumberOfColumns() == vector.vector_.size() && !vector.isRowVector_ &&
	"number of matrix columns must be equal to length of column vector");
	Vector resultVector;
	resultVector.vector_.resize(vector.vector_.size());
	
	for (unsigned row = 0; row < matrix.getNumberOfRows(); ++row)
		for (unsigned col = 0; col < matrix.getNumberOfColumns(); ++col)
			resultVector.vector_[row] += matrix(row,col)*vector.vector_[col];
	
	return resultVector;
}
\end{lstlisting}
\caption{Implementation of the matrix vector multiplication within the \texttt{Vector} class.}
\label{lst:matrix_vector_multiplication_implementation}
\end{codelisting}

Finally, we need to discretise the 1D heat equation in order to solve it using the CG method. This was a requirement we assumed before we started to implement the linear algebra library. A detailed derivation of the equation can be found in~\cite{Versteeg2007}, while the main steps are summarised below. The equation is given as

\begin{equation}
\Gamma \frac{\partial^2 T}{\partial x^2} = 0.
\label{eq:1D_heat_equation} 
\end{equation}

We discretise the equation using a finite volume approach with a central approximation for the temperature $ T $ across cell interfaces. This results in the discretised form of

\begin{equation}
\left(\frac{\Gamma}{\Delta x}\right)T^{n+1}_{i-1} + \left(2\frac{\Gamma}{\Delta x}\right)T^{n+1}_i + \left(\frac{\Gamma}{\Delta x}\right)T^{n+1}_{i+1} = 0.
\label{eq:1D_heat_equation_discretised}
\end{equation}

We may follow the notation in~\cite{Versteeg2007} and introduce the left and right neighbours of cell $ i $ as the west and east cells, while the current cell is abbreviated with the letter $ P $. This allows us to rewrite Eq.\cref{eq:1D_heat_equation_discretised} as

\begin{equation}
a_W T_W + a_P T_P + a_E T_E = 0,
\label{eq:simplified_discretisation}
\end{equation}

where $ a_W = a_E = \Gamma / \Delta x $ and $ a_P = a_W + a_E $. At the left and right boundary of the domain, we have to specify appropriate boundary conditions, which enter the equation as source terms $ s_P = -2\Gamma / \Delta x $ and $ s_U = 2\Gamma / \Delta x $. At the same time, we have to set either $ a_W $ or $ a_E $ to zero if it coincides with the boundary itself. In matrix form we get

\begin{equation}
\begin{bmatrix} 
a_P - s_P & a_E & \dots & & & & 0 \\
a_W & a_P & a_E &  & & & \\
\vdots & & & \ddots & & & \\
& & & & a_W & a_P & a_E \\
0 & & & & & a_W & a_P - s_P
\end{bmatrix}
\cdot
\begin{bmatrix}
T_1 \\
T_2 \\
\vdots \\
T_{n-1} \\
T_n
\end{bmatrix}
=
\begin{bmatrix}
s_U T_L \\
0 \\
\vdots \\
0 \\
s_U T_R
\end{bmatrix}.
\label{eq:discretised_equation_matrix}
\end{equation}

Here, $ T_L $ and $ T_R $ are the temperatures specified at the left and right boundary, respectively, resulting in a fully Dirichlet boundary problem. Solving the heat equation within a domain $ 0 < x < 1 $, as well as applying $ T_L=0 $ and $ T_R=1 $ as boundary condition with zero temperature as an initial condition, we obtain a linear profile for the temperature once the solution converges, which, given this test setup, allows for the temperature solution of the form $ T(x) = x $. This makes it particularly easy to ensure the final solution has converged towards the right solution. The system test is given in \cref{lst:system_test}.

\begin{codelisting}
\begin{lstlisting}
// skipping header includes, see full source code in online repository
int main() {
	const double gamma = 1.0;
	const unsigned numberOfCells = 100;
	const double domainLength = 1.0;
	const double boundaryValueLeft = 0.0;
	const double boundaryValueRight = 1.0;
	const double dx = domainLength / (numberOfCells);
	
	LinearAlgebra::Vector coordinateX(numberOfCells);
	LinearAlgebra::Vector temperature(numberOfCells);
	LinearAlgebra::Vector boundaryConditions(numberOfCells);	
	LinearAlgebra::Matrix coefficientMatrix;
	coefficientMatrix.setSize(numberOfCells, numberOfCells);
	
	// initialise arrays and set-up cell-centered 1D mesh
	for (unsigned i = 0; i < numberOfCells; ++i) {
		coordinateX(i) = i * dx + dx / 2.0;
		temperature(i) = 0.0;
		boundaryConditions(i) = 0.0;
	}
	
	// calculate individual matrix coefficients
	const double aE = gamma / dx;
	const double aW = gamma / dx;
	const double aP = aE + aW;
	const double sP = -2.0 * gamma / dx;
	const double sU = 2.0 * gamma / dx;
	
	// set individual matrix coefficients
	for (unsigned i = 0; i < numberOfCells; ++i)
		coefficientMatrix(i, i) = aP;	
	coefficientMatrix(0, 0) += -sP - aW;
	coefficientMatrix(numberOfCells - 1, numberOfCells - 1) += -sP - aE;
		
	for (unsigned i = 0; i < numberOfCells - 1; ++i) {
		coefficientMatrix(i, i + 1) = -aE;
		coefficientMatrix(i + 1, i) = -aW;
	}
	
	// set boundary conditions
	boundaryConditions(0) = sU * boundaryValueLeft;
	boundaryConditions(numberOfCells - 1) = sU * boundaryValueRight;
	
	// solve the linear system using the conjugate gradient method
	LinearAlgebra::ConjugateGradient CGSolver;
	CGSolver.setCoefficientMatrix(coefficientMatrix);
	CGSolver.setRHSVector(boundaryConditions);
	temperature = CGSolver.solve(1000, 1e-10);
	
	// calculate difference between obtain and expected solution
	LinearAlgebra::Vector difference(numberOfCells);
	for (unsigned i = 0; i < numberOfCells; ++i)
		difference(i) += std::fabs(temperature(i) - coordinateX(i));
	
	// ensure that temperature has converged to at least single precision
	assert(difference.getL2Norm() < 1e-8);
	
	return 0;
}
\end{lstlisting}
\caption{System test to verify the correct interactions between the vector, matrix and conjugate gradient class implementation.}
\label{lst:system_test}
\end{codelisting}

\begin{codelisting}
\begin{lstlisting}[language=]
Found ninja-1.10.0 at /usr/bin/ninja
[0/1] Running all tests.
1/10 unit test: set vector values                    OK           0.01s
2/10 unit test: scalar vector multiplication         OK           0.01s
3/10 unit test: vector vector multiplication         OK           0.01s
4/10 unit test: vector addition                      OK           0.01s
5/10 unit test: vector subtraction                   OK           0.00s
6/10 unit test: set matrix values                    OK           0.01s
7/10 unit test: scalar matrix multiplication         OK           0.01s
8/10 integration test: matrix vector multiplication  OK           0.01s
9/10 integration test: solve system using CG method  OK           0.01s
10/10 system test: solve 1D heat equation implicitly OK           0.12s

Ok:                 10  
Expected Fail:      0   
Fail:               0   
Unexpected Pass:    0   
Skipped:            0   
Timeout:            0   
\end{lstlisting}
\caption{Output after running \texttt{ninja -C build test} with all test unit, integration and system tests in the test suite.}
\label{lst:output_all_tests}
\end{codelisting}

\begin{figure}[t!]
	\centering
	\begin{subfigure}{\textwidth}
		\centering
		\includegraphics[width=\linewidth]{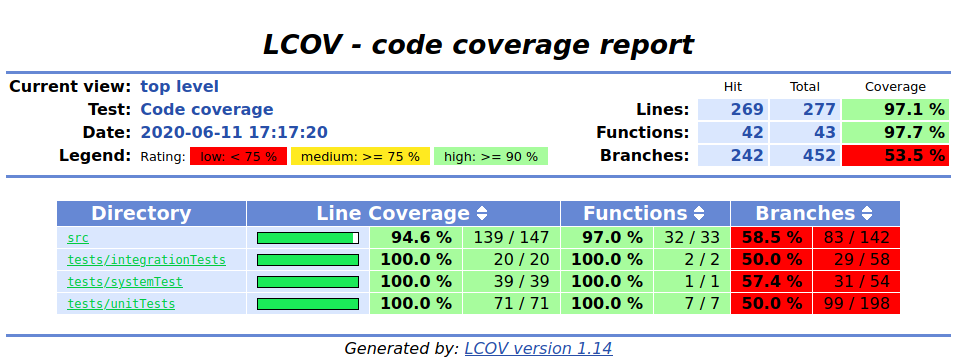}
		\caption{Coverage of the whole project, including the source and test sub-directory}
		\label{fig:lcov_all}
	\end{subfigure}
	\begin{subfigure}{\textwidth}
		\centering
		\includegraphics[width=\linewidth]{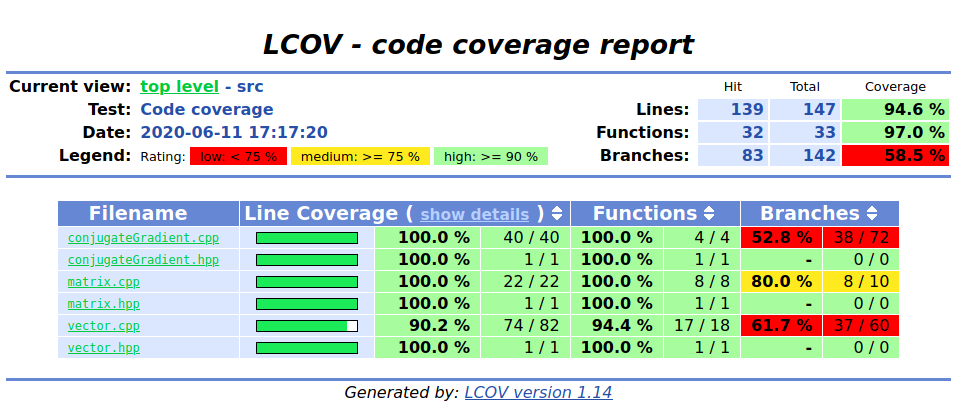}
		\caption{Detailed coverage of each file within the source sub-directory}
		\label{fig:lcov_src}
	\end{subfigure}
	\begin{subfigure}{\textwidth}
		\centering
		\includegraphics[width=\linewidth]{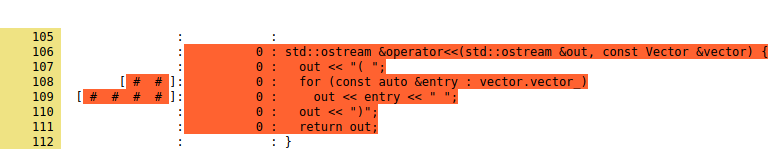}
		\caption{Example output of the vector class, here showing code that is not executed}
		\label{fig:lcov_vector}
	\end{subfigure}
	\caption{Coverage reports for the whole test project.}
	\label{fig:coverage_reports}
\end{figure}

With unit, integration and system tests in place, we may run the whole test suite and analyse their results. Invoking the command \texttt{ninja -C build test} results in the output shown in \cref{lst:output_all_tests} (wit added unit tests not further described here but available in the full project~\cite{Teschner2020}). As we can see, the whole test suite ran successfully which provides some confidence as to its correctness. Now that we know that all our tests have passed, we may also investigate how much of our code was actually tested. With coverage enabled, we can run \texttt{ninja -C build coverage} to automatically generated an HTML coverage report, which will be stored within the build sub-directory. A sample output of this is shown in \cref{fig:coverage_reports}. In \cref{fig:lcov_all}, we can see the output for the whole project, which includes the test and source sub-directories. In our case, however, we are not interested in the coverage of the test file but rather those located in the source sub-directory. However, in this project overview we can already see that all of our written tests cover 94.6\% of the code within the source sub-directory. \Cref{fig:lcov_src} provides an overview of the files within the source directory and we can see that except for the \texttt{vector.cpp} file, we have 100\% test coverage. For a small project like this one, it is relatively simple to achieve such high coverage metrics, however, with an increasing project size that metric is likely to get lower. It is not important to aim for full test coverage here, rather the core business logic should be tested. For the same reason, we can see that we have a rather low branch coverage (branches are encountered whenever the code can split, for example in if statements) which are, however, not a problem as we have ensured that our use case (here the simulation of a 1D heat equation) is working. Trying to increase the test coverage for the sake of reaching a certain threshold goes against best practices~\cite{Khorikov2020} which results in development time spent on writing unnecessary tests which would not change the outcome of the simulation of the 1D heat equation (it has already passed).\\
In \cref{fig:lcov_vector}, we inspect the \texttt{vector.cpp} file to see which part has not been covered by tests. We see here a function that overloads the output stream operator that allows us to write our vector class to screen. This feature was implemented for debugging but is not required in the production code. Thus, we have two options here, either we remove the code (in line with extreme programming rules to only produce code which is absolutely necessary) and obtain a higher test coverage, or, we leave the code but ignore it during our tests. The latter approach is chosen, as it is a useful feature to have, while it is OK to ignore the code during testing, as the 1D heat equation system test will pass regardless of this functionality.\\
\\
Finally, there are a few comments in order. First of all, the purpose of this example was to show how unit, integration and system tests can be integrated into a scientific and academic development cycle. As such, this case study is of educational purpose only and therefore readability has been traded for performance. In a real production code, for example, the matrix should not be stored with $ N^2 $ entries, as most of them are zeros. In this case, we would typically resort to special matrix storage systems, such as the compressed row storage (CRS). Furthermore, we said that for scientific and academic software developments, user and system testing become a single unit, making the system test given above a user test at the same time. We have also not specified the end-user application. We only specified that it should be able to solve the 1D heat equation. Assuming that this is part of a larger thermal analysis software, for example, we would probably provide more abstract interfaces to the system test, for example, by defining a scalar field class (used as an abstraction for the temperature) that uses the vector class as a storage container while providing the access logic to the vector class, such as defining iterators. These are specified as controllers by Khorikov~\cite{Khorikov2020} and ensure that we split the algorithm implementation from the access logic, following the single responsibility principle. We would probably also define operators (such as a Laplacian operator) that would automatically produce the matrix coefficients based on the assumed temperature profile between cells and define a computational matrix class, using the already existing matrix class as a container while handling the set-up based on the operators internally. We have also opted to include the system and integration tests within the test suite execution, which for larger projects becomes time intensive and thus they should be removed and tested less frequently. For small projects, however, it may still be acceptable to execute all tests at the same time. While the focus here was on minimal overhead with as few tools as possible, it should be highlighted here that for larger projects, it would be advisable to make use of an external testing library. Popular choices in c++ are, for example, \texttt{Boost Test Library, CppUnit, Google Test} and \texttt{QtTest}. However, making use of the above would have only distracted from the core objective here, which was to highlight the use of unit, integration and system tests. Only a few code examples have been shown here, while the whole project may be accessed online~\cite{Teschner2020}.

\section{Conclusion}
\label{sec:conclusion}

Scientific software development is often used to test newly developed models or to gather data, from which new knowledge is derived. It is mandatory to ensure the correctness of the software during production but also once it is extended. Test-drive development (TDD) is a common approach adopted in agile software development. Agile development methodologies like Scrum define various roles, some of which cannot be occupied by the same person, which makes it impractical for most scientific software projects with is a single developer which is also the end-user at the same time. Furthermore, Scrum assumes that work can be conducted in sprints which may not be feasible in an academic environment, as time cannot be allocated solely to code development. However, there are many good practices that can be adopted from an agile development perspective and this study focused on how TDD can be adopted within an academic environment. Within TDD, unit, integration and system tests are produced, which in the classical testing matrix are extended by user and release tests. For small scientific and academic software developments, however, all user testing can be removed and merged with the system tests, which can in some cases be automated as shown here in \cref{sec:case_study_developing_a_testing_suite_for_a_linear_algebra_library}. More generally speaking, however, system tests (and user tests) require manual checking and it may be necessary to automate the system tests as much as possible (for example, having an automated html report generated from results) which can then be inspected by a set of trained eyes. The developed test suite, consisting of unit, integration and system test can be used to continuously check that newly added code does not break existing code. This is usually done by a continuous integration (CI) server, which runs the test suite every time a new piece of code is submitted to the server. For smaller, single-developer, projects, however, especially those where a piece of code is generated to test a new hypothesis, it may seem excessive to set up a continuous integration web server or pay for third-party hosting. Those codes are developed once and it is not uncommon for them to not be used again, and therefore continuous integration is still a good idea in principle, but may be done directly on the developer's machine instead of outsourcing it to a server. Both TDD and CI can thus be reduced to a set of local tasks which may be completed entirely offline. In order to achieve this task, a practical guide was provided in \cref{sec:software_testing}, showing how a sample project could be set up using the meson build system. Meson provides a relatively simple syntax (based on python) to setup the build targets while providing a simple testing interface which was used to create the test suite. While c++ was used here to set up the project, meson has support for many more languages. The build scripts are always written in the same way and we only need to specify the language used once so that meson can deal with all the required compilation specifics for us. This sample project was then used in \cref{sec:case_study_developing_a_testing_suite_for_a_linear_algebra_library} to create a simple linear algebra library that was used to solve a 1D heat equation as a system test. Furthermore, the procedure of the TDD cycle was demonstrated creating unit, integration and system tests. The fully developed project can be found online~\cite{Teschner2020}, which may serve as a starter project for future software development purposes. The developed project and annotations given herein are hoped to show that TDD and CI can be achieved relatively easily for even small scientific software developments. The advantage is obvious; with an increased test coverage and quality tests, we can ensure that our software is working as intended and that software bugs are not by mistake overlooked. The reason for this article then is simple; the literature review has shown that outside of industrial software development projects, there is seldomly a strict testing policy in place. It has also shown that those who make rigorous use of testing regard test code just as valuable as production code and that the increased time spent on writing test is offset by the reduced time required for debugging. Using the practical guide developed within this article will help to increase software quality and prevent deriving wrong or inconclusive knowledge from generated data due to undiscovered software defects. It is hoped that the reader can appreciate the value tests are adding to software project and that systematic testing will be favoured in the future over the rather outdated ad-hoc testing approach.

\bibliographystyle{siamplain}
\bibliography{references}
\end{document}